\documentclass{article}
\usepackage{fixltx2e}
\usepackage[latin1]{inputenc}
\usepackage{amsmath,amsfonts,amssymb,bm}
\usepackage{graphicx}
\usepackage{cite}
\usepackage[tight,hang,small,bf]{subfigure}

\begin{document}
\title{100 MHz amplitude and polarization modulated optical source
for free-space quantum key distribution at
850 nm}

\author{M.~Jofre,~A.~Gardelein,~G.~Anzolin,~G.~Molina-Terriza,\\~J.P.~Torres,~M.W.~Mitchell~and~V.~Pruneri}

\date{}

\maketitle

\begin{abstract}
We report on an integrated photonic transmitter of up to 100 MHz repetition
rate, which emits pulses centered at 850 nm with arbitrary
amplitude and polarization. The source is suitable for free space quantum key distribution applications. The whole transmitter,
with the optical and electronic components integrated, has reduced
size and power consumption. In addition, the
optoelectronic components forming the transmitter can be
space-qualified, making it suitable for satellite and future space
missions.
\end{abstract}


\section{Introduction}
In many applications, \textit{free space optical} (FSO) communications
is the technology of choice to transmit information, especially when
fiber optical cabling is not easily achievable or its installation is
too expensive \cite{Carbonneau1998}. Compared to \textit{radio frequency} (RF) techniques, its main
advantages lie in high data rates (up to several Gb/s), minimum free space
losses due to the small optical beam divergence and absence of regulatory
issues thanks to the low interference level \cite{Garlington2005,O'Brien2003,Davis2003}.
Therefore FSO communication is favorable for high data-rate, long-range
point-to-point links, where the terminal size, mass, and power
consumption are subjected to strong limitations, such is the case
of aeronautical or space platforms.

An important issue in today's information society is the security
of data transmission against potential intruders, which always put
at risk the confidentiality. Current methods to increase security
require that two parties wishing to transmit information securely
need to exchange or share one or more keys. Once the key has
been exchanged, the information can be transferred in a provable
secure way using a one-time pad. Therefore, the security of
the information transmission is based exclusively on the security
of the key exchange. Quantum cryptography, or more precisely
\textit{Quantum Key Distribution} (QKD), guarantees absolutely
secure key distribution based on the principles of quantum
physics, since it is not possible
to measure or reproduce a state (eg. polarization or phase of a
photon) without being detected \cite{RevModPhys.77.1225}. The key is
generated out from the measurement of the
information encoded into specific quantum states of a photon. In
particular, if a two-dimensional quantum system is used,
information is said to be encoded into qubits. For example, a
qubit can be created using properties such as the polarization or the phase of a photon.

The first QKD scheme, due to Bennett and Brassard \cite{Bennett1984,Bennett1992}, employs single
photons sent through a quantum channel, plus classical communications
over a public channel to generate a secure shared key. This scheme is
commonly known as the BB84 protocol. Although single photon sources may
be very useful for quantum computing, they are not strictly
required for QKD. This, and the relative difficulty of generating
true single photons, motivates new approaches based
on conventional light sources \cite{scarani-2008}. Indeed,
attenuated laser pulses or \textit{faint pulse sources} (FPS),
which in average emit less than one photon per pulse, are often
used as signals in practical QKD devices. The performance
limitations of attenuated pulse systems had initially led to
believe that single photon sources would be indispensable for
building efficient QKD systems. However, the introduction of the
decoy-state protocol
\cite{PhysRevLett.94.230504,PhysRevA.72.012326} made possible a
much tighter bound for the key generation rate, achieving an
almost linear dependency of the latter on the channel
transmittance. In this way, the technologically much simpler faint
pulse systems can offer comparable QKD security with respect to
single photon sources. Another key feature of QKD is that the
security is linked to the one-time-pad transmission, i.e. the key
has to be used once and has to be equal or similar in size to the
information being transmitted. It is thus evident the importance
of developing faint pulse sources and systems for QKD which can
generate high key bit rates. The highest Secure Key Rate reported to date over 20 Km of optical telecom fiber is of 1.02 Mb/s \cite{Dixon:08} and 14.1 b/s over 200 Km \cite{Chen2009}, while the achieved speed over 144 Km free space link is of 12.8 Kb/s \cite{Schmitt-Manderbach2007} and 50 Kb/s over 480 m \cite{Weier2006}.

The goal of QKD is to allow to distant parties to share a common
key in the presence of an eavesdropper. Therefore, the most
important question of QKD is its security. Therefore, an important
aim of this work is to demonstrate a system to generate pulses
that differ only in polarization, while being indistinguishable in
the other degrees of freedom that characterize the quantum state
of photons, such as arrival time, optical frequency, and spatial
mode. In other words, to generate pulses which contain no
side-channel information correlated to the polarization. We note
that previous implementations based on multiple lasers
\cite{Schmitt-Manderbach2007,Weier2006,Kurtsiefer,4528718} have
attempted to achieve time-frequency indistinguishability by laser
pre-selection, current and temperature adjustment, and temporal
and spectral measurements. Apart from being expensive and
cumbersome, this kind of tuning has limited stability due to the
inevitable aging of laser diodes. It is worth noting that the
temporal and spectral distributions reported to date indicate
indistinguishability in the time and frequency bases, but leave
open the question of distinguishability based on other pulse
characteristics such as chirp.

A related issue which arises in a decoy-state protocol is possible
side-channel information indicating the pulse intensity. Intensity
level modulation could be achieved rapidly and conveniently
by modulating the laser current. This method of modulation,
however, induces strong nonlinearities and causes strong phase
modulation, which makes it difficult to control the temporal and
spectral shape of the output pulses.

In this paper we report the development of a novel integrated
pulse source which can reach rates as high as 100 Mb/s at 850nm
modulated in amplitude and polarization. For QKD applications, it
has been simulated that the source could achieve a Secure Key Rate
of the order of 500 Kb/s at 20 Km using decoy-state protocol. The
source is capable to generate pulses at around 850 nm with at
least three different intensity levels (i.e. number of photons per
pulse) and four different polarization states. The proposed FPS ensures indistinguishability among
the different intensity and polarization
pulses and ensures phase incoherence of consecutive generated states. It
is based on a single diode emitting a continuous pulse train
externally modulated in amplitude and polarization. The
wavelength, reduced power consumption, compactness and space
qualifiable optoelectronic components constituting the source make
it very suitable for space transmission, for free space quantum
and classical communication links. One of the foreseen
applications is its use to overcome the distance limit of QKD in
optical fibers \cite{Chen2009,Takesue2007}, by creating a global
security network among very distant places on earth through
satellite communication
\cite{PhysRevLett.91.057901,Rarity2002,Bonato2009}.

\section{The integrated faint pulse source}\label{ExperimentalSetup}
\begin{figure}[htp]
\centering
\includegraphics[width=\columnwidth]{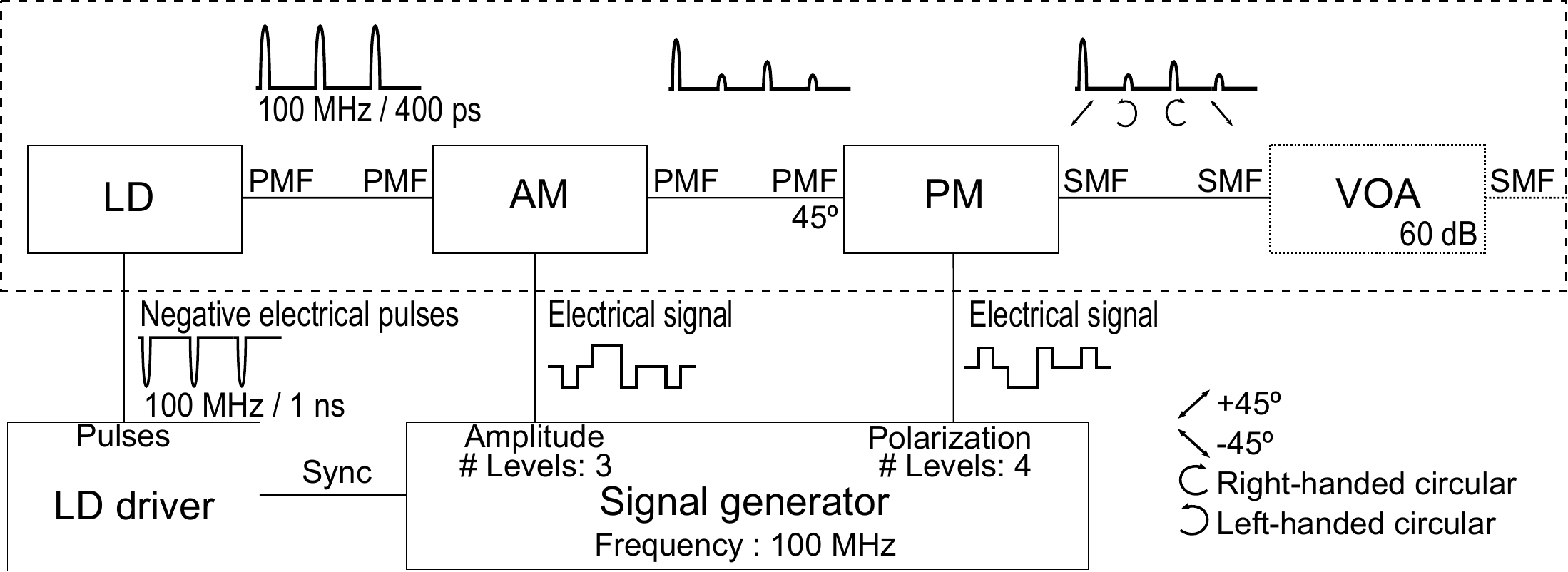}
\caption{Schematic of the QKD source. [LD] denotes a laser diode, [AM] an amplitude modulator, [PM] a polarization modulator and [VOA] a variable optical attenuator.}
\label{Fig:FPS-IMPM_Blocks}
\end{figure}

In order to use it for space applications, the proposed integrated
FPS source for FSO communication consists of commercially
available space-qualified discrete components; single
semiconductor laser diode emitting a continuous pulse train at 100
MHz followed by integrated (waveguide) amplitude and polarization
lithium niobate (LiNbO$_{3}$) modulators (Figure
\ref{Fig:FPS-IMPM_Blocks}).

A \textit{distributed feedback} (DFB) \textit{laser diode} (LD) at around 850 nm is driven at 100 MHz train of electrical pulses. The optical pulse of about $400$ps is generated via a current pulse of about $1$ns duration. In fact, the laser is biased using a DC current of $24$mA, far below threshold $36$mA, and it is directly modulated using a strong RF current of $50$mA (peak value) so that the optical pulse is generated \cite{Petermann1991}. The generated optical pulses do not have any phase coherence among them due to the fact that the laser is set below and above threshold from one pulse to the subsequent one, thus producing a random phase for each pulse. The output mirror reflectivity ($R$) of the DFB structure is $30$\%, the cavity length ($L_c$) $300\mu$m and the active medium refractive index ($n$) $3.6$ \cite{Sadovnikov1995,Li1998}. The \textit{round-trip-time} (RTT) given by $2L_c n/c_0$, where $c_0$ is the speed of light in vacuum, is $\approx 20$ps. In one pulse train period ($10$ns) the optical pulse power left in the laser cavity, after going below threshold, has bounced back and forth $\approx 500$ times. Between pulses, the laser is biased at only $66$\% of threshold, so that transmission loss through the output mirror is much greater than the round-trip gain. Conservatively assuming a round-trip loss $\ge 1$dB, the $\ge 500$dB loss from $500$ passes will atenuate any coherence to a very low level. At the same time, incoherent spontaneous emission is generated, further obscuring any possible coherence between optical pulses \cite{Petermann1991}. In terms of partial coherence theory, it is expected a first-order degree of coherence $g^{(1)}(\tau)$ which drops rapidly to zero for $\tau$ larger than the pulse duration. This is consistent with the observed spectrum, which implies a coherence length of order $0.75$ m {\em during} the pulse. Note that the spectrophotometer response is produced by photons coming from the pulses and the coherence length of about $0.75$ m which can be calculated from the measured bandwidth, as it was explained above, decreases significantly when the pulse extinguishes, thus making the inter-pulses value much smaller than the distance between consecutive pulses (about $3$ m).

In this way, phase incoherence of consecutive generated states,
which otherwise would be detrimental for the link security, is
achieved. Then, the pulse train is sent through a
\textit{polarization maintaining fiber} (PMF) into an \textit{amplitude modulator} (AM)
(eg. a Mach-Zehnder modulator in LiNbO$_{3}$) that will randomly
generate the three different levels of intensity. Note that if the DFB laser diode were
driven in pure \textit{continuous wave} (CW) mode (no
pulse train) and externally modulated to generate the pulses, two
potential issues would occur: (i) pulses with different energies
(number of photons) would unavoidably have different temporal and
spectral shapes due to the nonlinear electro-optic response
(optical output as a function of driving voltage) of the amplitude
modulator; (ii) there would be phase coherence between the pulses
due to the relatively long coherence time (narrow spectrum) of a
DFB structure, thus increasing the vulnerability of the QKD
transmission \cite{lo-2007-8}.

After the AM, the pulses are injected into
a \textit{polarization modulator} (PM) through a PMF. The polarization modulator
is in fact a waveguide LiNbO$_{3}$ phase modulator where the PMF
input axis is oriented at 45\textdegree with respect to the
optical axis. In this way, the two orthogonal equal amplitude
polarization components of the electromagnetic field that
propagates in the crystal experience a refractive index
difference, which is proportional to the voltage applied to the
modulator. By applying different voltages one can thus change the
state of the output polarization, in particular linear
+45\textdegree, -45\textdegree, right-handed circular and
left-handed circular. The optical pulses present a spectrum within
the acceptance bandwidth of the two modulators, so that amplitude
and polarization modulation can be achieved with high extinction
ratio.

A proper electronic control of the different intensities and polarization states generated, at Alice, for the different states is fundamental in order to perform a QKD transmission. The synchronization and setting of the different optical components of the source is implemented by an automatic control which is split into two working operations. The control system first synchronizes and calibrates the driving signals timings and amplitudes to the AM and PM, and secondly generates the appropriate driving signals for the BB84+Decoy protocol transmission.

For the implementation of a QKD system using decoy-state
protocol, besides four different polarization states, the FPS
source should generate three intensity levels (optimally 1/2,
1/8 and 0 photons in average per pulse \cite{PhysRevA.72.012326}) using a \textit{variable
optical attenuator} (VOA) in order to operate in the single photon
regime. The optical pulse duration $\approx 400$ps and the pulse peak power $3.5$mW which corresponds to $1.4$pJ energy per pulse, thus a number of photons per pulse $\approx 6\cdot 10^{6}$. In order to get a mean photon number for the signal state which is within an optimum range for the distances of interest \cite{PhysRevA.72.012326}, the VOA has to introduce an optical attenuation of $\approx 70$dB.

The chosen FPS wavelength (850 nm) is optimum for free
space operation considering attenuation (due to scattering,
absorption and diffraction) and single-photon detector's quantum
efficiency \cite{scarani-2008}.

\newcommand{\polsym}{{p}}
\newcommand{\vecpolsym}{{\bf p}}
\newcommand{\ampsym}{{\cal E}}
\newcommand{\scalesym}{{\eta}}
\section{Description of generated states}\label{Sec:Description_generated_states}
In a BB84 protocol scheme implementing the decoy-state protocol
different pulses should differ in polarization and amplitude while remaining indistinguishable
in other characteristics, including temporal shape and frequency spectrum. If the pulses differ in spectrum, for example, an eavesdropper could use spectral measurements to infer the sent polarization without actually measuring it. Removal of this kind of {\em side-channel} information is thus critical to the security of the protocol. Since the information is encoded in the polarization state, the statistical similarity between pulses of different polarizations but same intensity level is more relevant than that of different intensity level but same polarization to prevent information leakage from the quantum
link. Here we consider the quantum optics of side-channel information, limiting the discussion to pure states and simple measurements. A full treatment including mixed states and generalized measurements will be the subject of a future publication.

We consider a source that produces pulses with amplitudes $\ampsym_l$, polarizations $\vecpolsym_l$ and pulse shapes $\Pi_l(t)$. Without loss of generality we assume the polarizations and pulses shapes are normalized $\vecpolsym^*_l \cdot \vecpolsym_l = \int dt \, \Pi_l^*(t) \Pi_l(t) = 1$. In a classical description, the field envelopes are
\begin{equation}
\mathbf{E}_{l}(t)=\ampsym_l \vecpolsym_{l}\Pi_{l}(t)
\label{Eq:ClassicalPulse}
\end{equation}
The corresponding quantum state is a generalized coherent state 
\begin{equation}
\left|{\bf \alpha}_{l}\right> \equiv D_l(\scalesym \ampsym_l \vecpolsym_{l}) \left| 0 \right>
\end{equation}
where $ \left| 0 \right>$ is the vacuum state and $D_l({\bf x}) \equiv \exp[{\bf x}\cdot {\bf A}_l^\dagger - {\bf x}^*\cdot {\bf A}_l ]$ is a displacement operator, defined in terms of the mode operator ${\bf A}_{l} \equiv \int dt\,\Pi_{l}^*(t){\bf a}(t) $, ${\bf a} \equiv (a_x,a_y)$ is a vector of annihilation operators, with $[a_p(t),a_q^\dagger(t')] = \delta(t-t')\delta_{p,q}$ for $p,q \in \{x,y\}$. A scaling factor $\scalesym$ is included to convert from photon units to field units, chosen such that the positive-frequency part of the quantized electric field is $\hat{\mathbf{E}}(t) = \scalesym^{-1} {\bf a}(t)$. It is easy to check that $\left<\alpha_l\right| {\bf a}(t) \left|\alpha_l\right> = \scalesym \ampsym_l \vecpolsym_{l}\Pi_{l}(t)$, so that the average quantum field$\left<\alpha_l\right| \hat{\mathbf{E}}(t)  \left|\alpha_l\right> =\ampsym_l \vecpolsym_{l}\Pi_{l}(t)$ in agreement with Equation \ref{Eq:ClassicalPulse}.

Quantum mechanics allows measurements on the pulse-shape $\Pi$ without measurement of the polarization $\vecpolsym$. For example, the number operator $N_l \equiv {\bf A}^\dagger_l \cdot {\bf A}_l = A_{l,x}^\dagger A_{l,x}+A_{l,y}^\dagger A_{l,y}$ counts photons in the mode $\Pi_l$ independent of $\vecpolsym_l$. If the modes $\left\{\Pi_l\right\}$ are different, an eavesdropper could use state-discrimination techniques \cite{Bergou2004,Barnett09} to determine $l$ (and thus the secret key) {\em without} disturbing $\vecpolsym$. This kind of eavesdropping would not be detected by Bob's polarization measurements. For this reason, it is critical to guarantee that this kind of {\em side channel} information is not present in the sent optical pulses. The similarity between the various $\Pi_l$ can be quantified by an overlap integral: $[{A}_{l,p},{A}_{m,q}^\dagger]  = \int dt \Pi_l^*(t) \Pi_m(t)[a_p,a^\dagger_q] \equiv S_{lm} \delta_{p,q}$, so that for example two states with equal amplitudes $|\ampsym_l|=|\ampsym_m|$, $\left<\alpha_m\right| N_l \left|\alpha_m\right>/\left<\alpha_l\right| N_l \left|\alpha_l\right> = |S_{lm}|^2$. Finally, we note that it is possible for pulses to have the same spectra and temporal shape but still be distinguishable, for example if they have different chirp. For this reason, establishing that two (or more) distinct sources produce indistinguishable pulses is not easy.

Our strategy to eliminate side-channel information in the pulse shapes is to dissociate pulse generation from the setting of polarization and amplitude levels. As described in the previous section the FPS consists of a single laser diode emitting a continuous train of optical pulses followed
by an AM, a PM and a VOA. Considering that the laser operation is the same for each pulse sent, 
and that both the AM and PM control voltages are held constant over the duration of the pulse, we can assume that the pulse shape does not depend on the sent amplitude and polarization. This assumption is confirmed by measurements shown in Section \ref{Experimental_measurements}.
The complex expression of the pulsed electromagnetic field exiting the FPS can
be written as
\begin{equation}
\mathbf{E}(t)=\sum_i A \alpha_i e^{j\phi_i}e^{j\beta_i}\frac{\bf{\hat x}+e^{j\gamma_{i}}\bf{\hat y}}{\sqrt{2}}\Pi\left({t-iT}\right)
\end{equation}
where $t$ is the time, $T$ is the pulse train period and $A,\phi_i,\Pi$ are the amplitude, phase, and shape, respectively, of the optical pulse generated by the
LD. $\alpha_i, \beta_i$ describe the transmission and introduced phase, respectively, of the AM.
$\gamma_{i}$ is the phase difference between $\bf{\hat
x}$ and $\bf{\hat y}$ introduced by the polarization modulator in
order to generate the different polarization states. 

Another security consideration is optical coherence between successive pulses, which could in principle be used
for eavesdropping attacks \cite{lo-2007-8}. As the LD is taken below threshold between pulses, each new pulse will start up from vacuum fluctuations, and will have a random overall phase $\phi_i$, thus eliminating coherence between successive pulses and thus among states. Similarly, any information contained in the AM phase $\beta_i$ is washed out by the random $\phi_i$. 

\section{Experimental measurements}\label{Experimental_measurements}
\begin{figure}[htp]
\begin{center}
\subfigure[]{\includegraphics[angle=0,width=0.8\columnwidth]{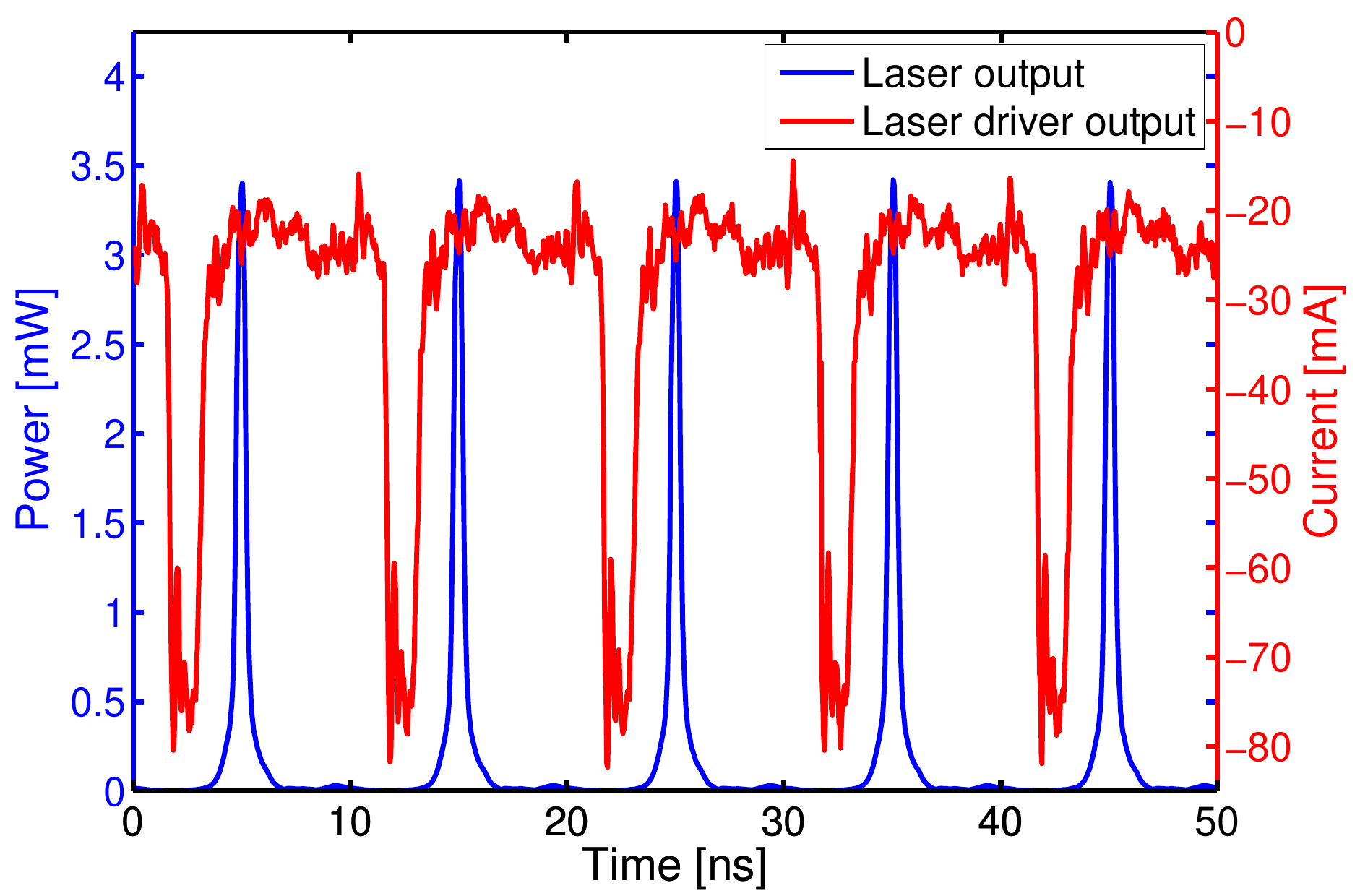}}
\subfigure[]{\label{Fig:Laser_output_and_Laser_driver_output_time_distribution}\includegraphics[angle=0,width=0.8\columnwidth]{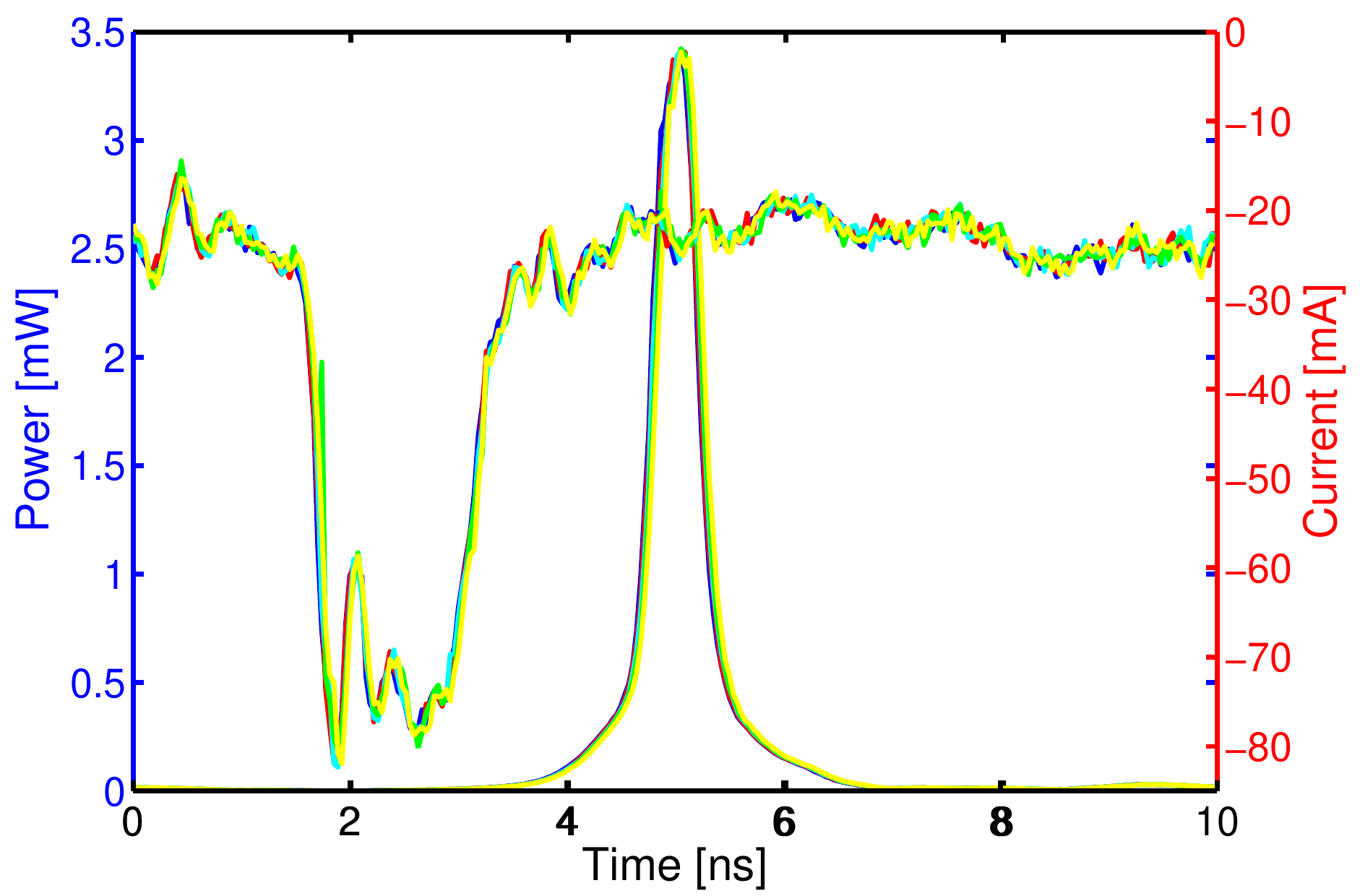}}
\caption{Laser diode output and laser driver output results. (a) Generated cw train of optical pulses at 100MHz experimental results. The optical pulses at 100MHz are generated using a DFB laser diode (upward pulses) directly modulated using the driving RF electrical pulses (negative pulses). (b) Time distribution of five pulses from the laser diode and the laser driver.}
\label{Fig:Laser_output_and_Laser_driver_output}
\end{center}
\end{figure}

Figure \ref{Fig:Laser_output_and_Laser_driver_output} (a) shows the
train of optical pulses generated by the laser diode when driven
by electrical pulses of 1 ns at 100 MHz. The resulting optical
pulse duration is about 400 ps. Since the obtained cw train of
optical pulses are all generated in the same way, they can be
assumed to be indistinguishable thus having no side-channel
information. Furthermore, the short optical pulse duration of 400
ps (small duty cycle) has the advantage to increase the signal to
noise ratio since the measurement window (detection time) in the
receiver can be reduced. The DFB laser diode is driven in direct modulation with a strong RF driving signal with $24$mA DC bias current, far below threshold $36$mA, thus producing highly similar optical pulses and jitter as low as $100$ps, rise time $65$ps and fall time $129$ps, as shown in Figure \ref{Fig:Laser_output_and_Laser_driver_output} (b). From Figure \ref{Fig:Laser_output_and_Laser_driver_output} (b) one can see that the ringing of the laser driver current is repeatable from pulse to pulse, thus producing the overlapped temporal profile of several optical pulses, captured in real-time. The traces are indistinguishable by eye, indicating a very small pulse-to-pulse variation of energy, timing, and wave-form. Furthermore, the optical pulse bandwidth
is small enough to enter the acceptance bandwidth of the subsequent polarization modulator.

\begin{figure}[htp]
\begin{center}
\subfigure[]{\includegraphics[angle=0,width=0.64\columnwidth]{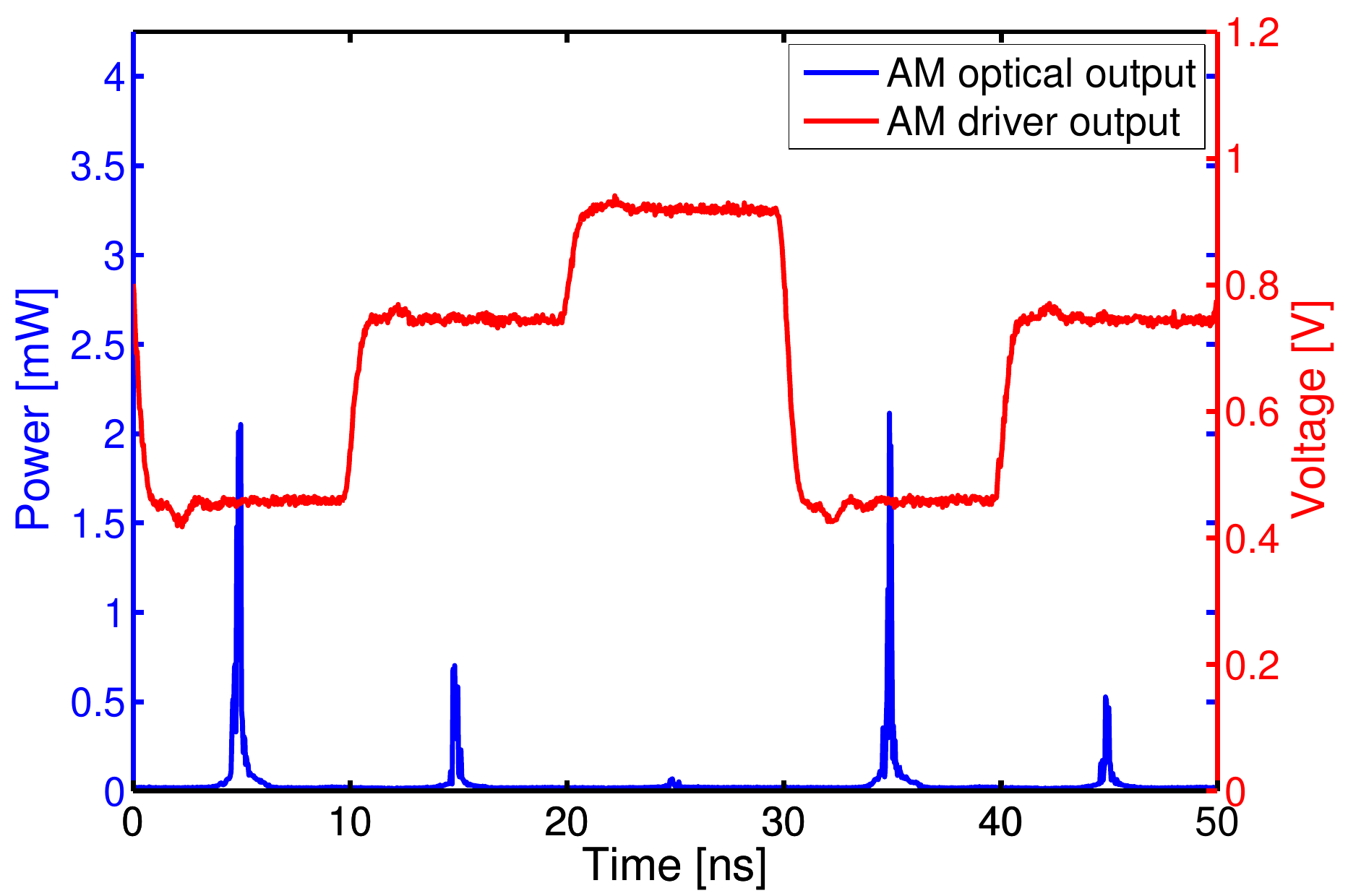}}
\subfigure[]{\includegraphics[angle=0,width=0.34\columnwidth]{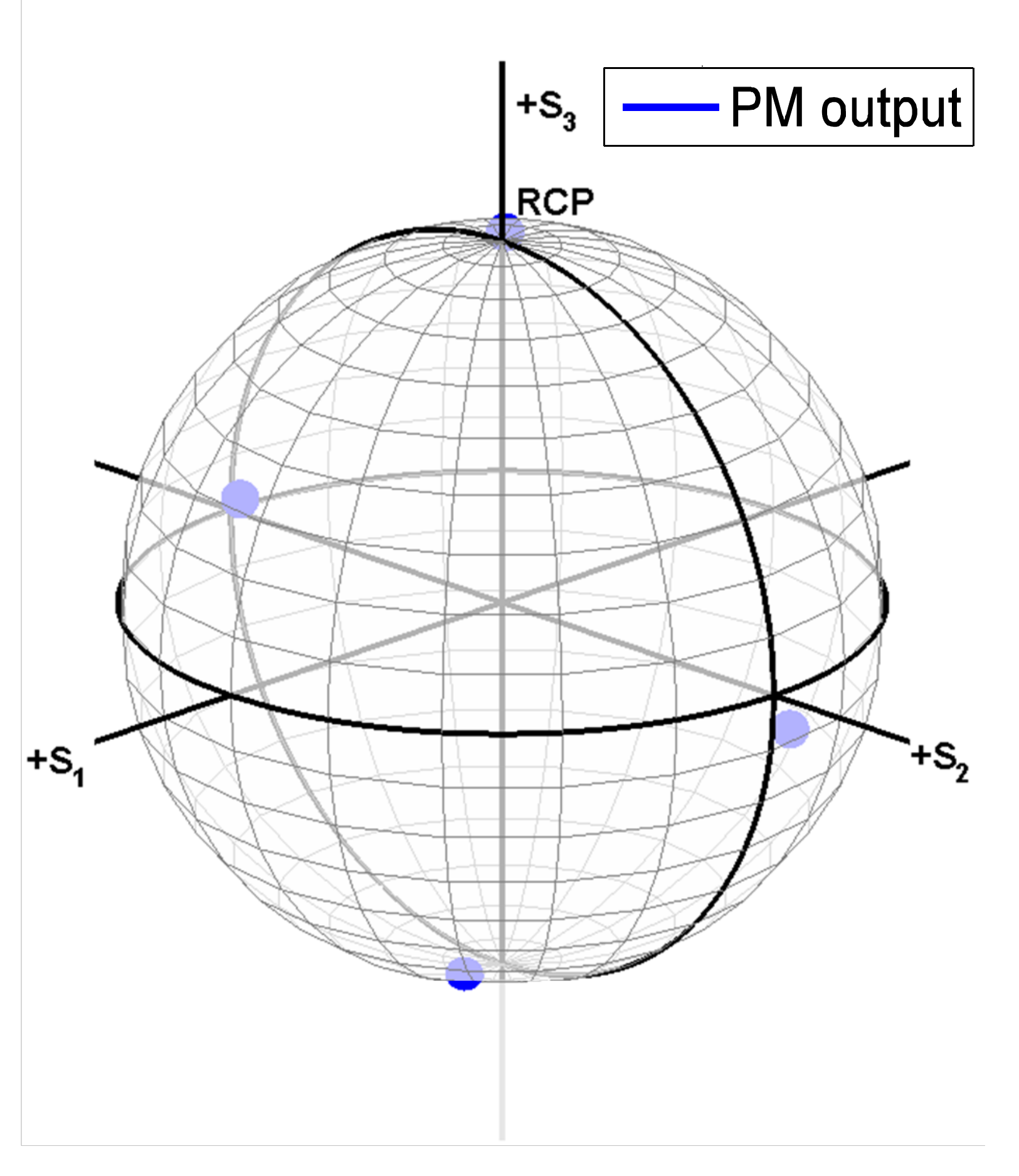}}
\caption{Amplitude and polarization modulators experimental results. (a) Amplitude modulator experimental results. Different intensity levels are generated at 100 MHz, where the modulator window is much larger than the pulse time width. (b) Different polarization states are generated (shown on the Poincare sphere), in particular it is shown four polarization states: +45\textdegree, -45\textdegree, right-handed circular and
left-handed circular; sufficient to implement a BB84 protocol.}
\label{Fig:AM_3_levels_and_PM_Poincare}
\end{center}
\end{figure}

\begin{figure}[htp]
\begin{center}
\subfigure[]{\includegraphics[angle=0,width=0.49\columnwidth]{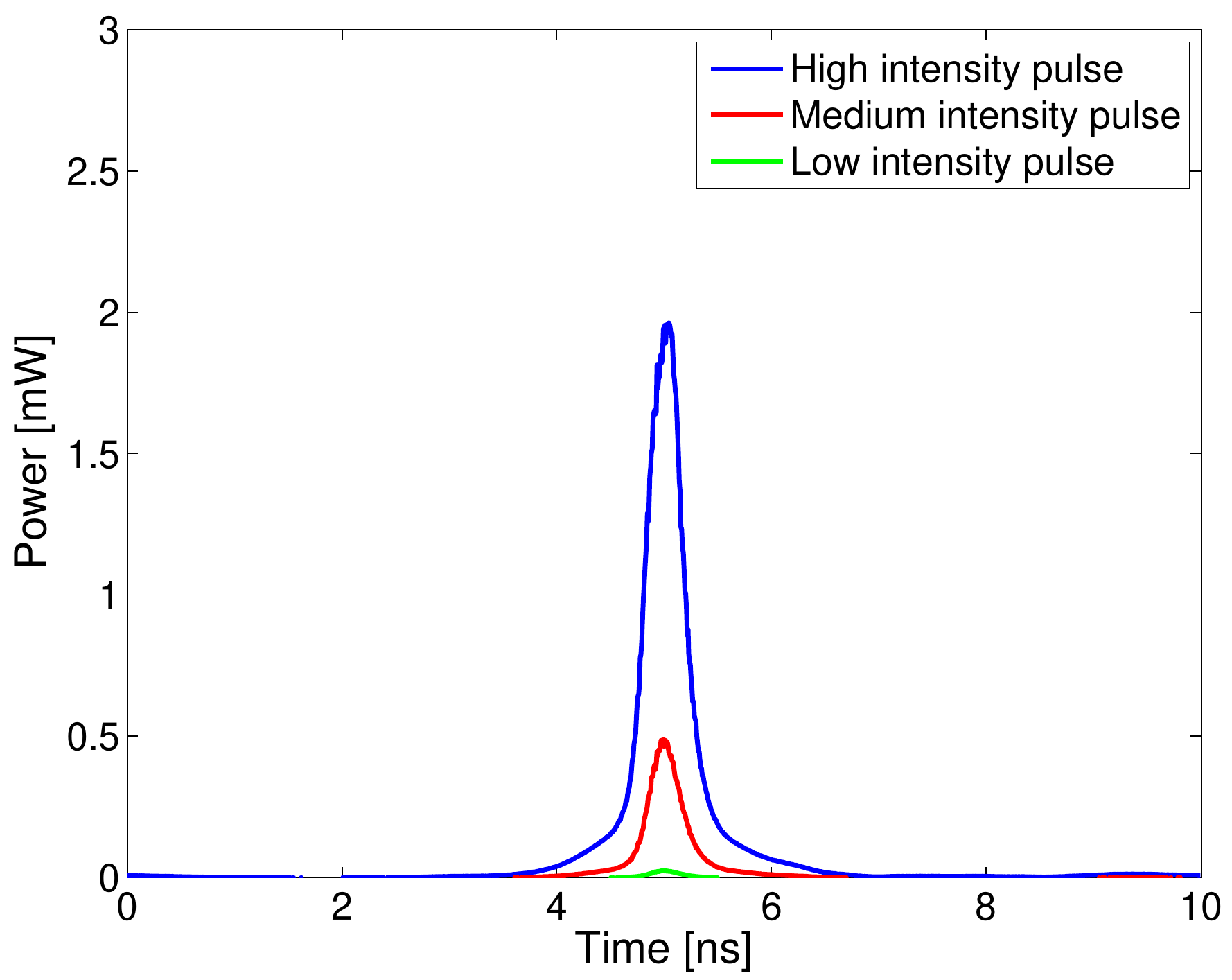}}
\subfigure[]{\includegraphics[angle=0,width=0.49\columnwidth]{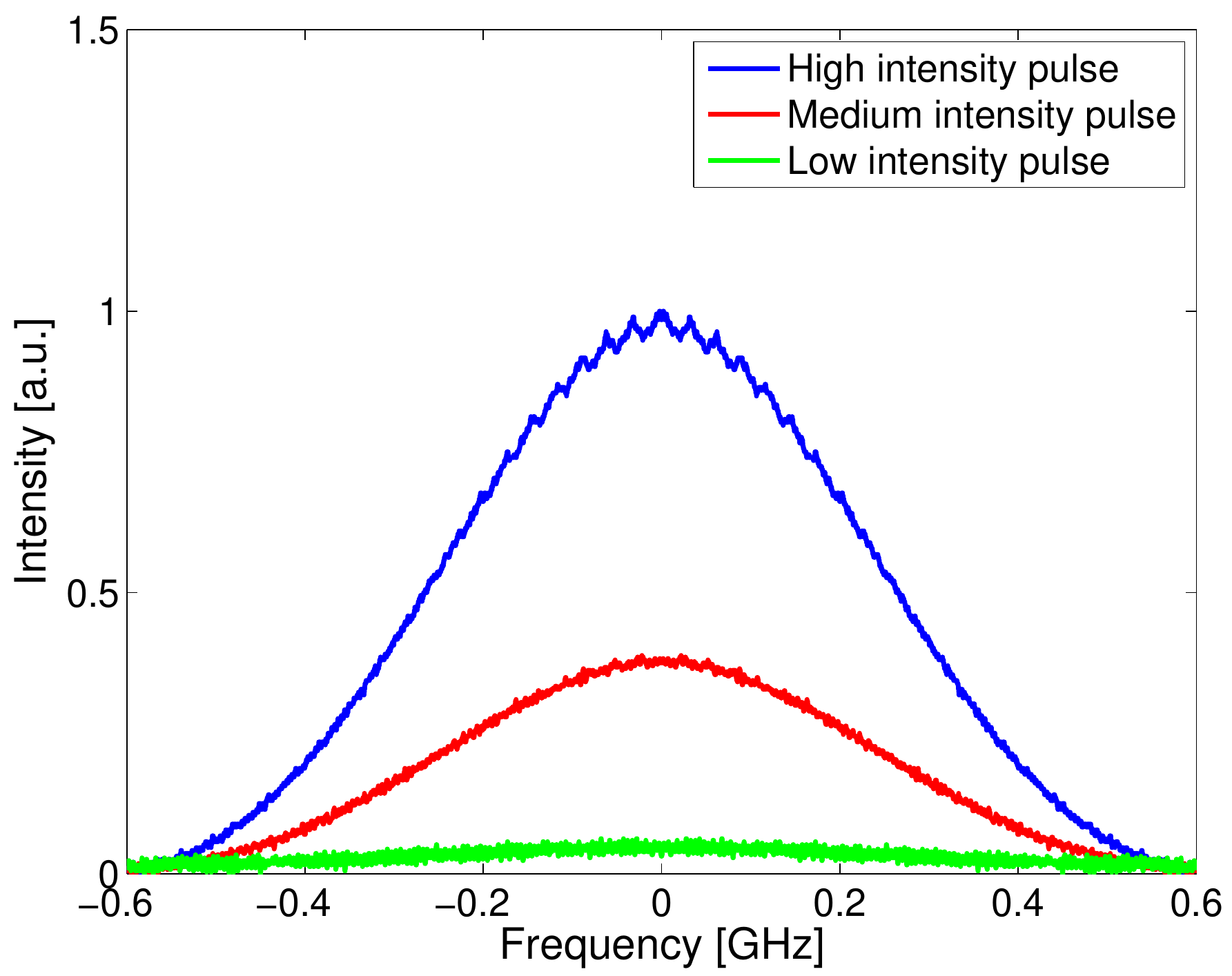}}
\subfigure[]{\includegraphics[angle=0,width=0.49\columnwidth]{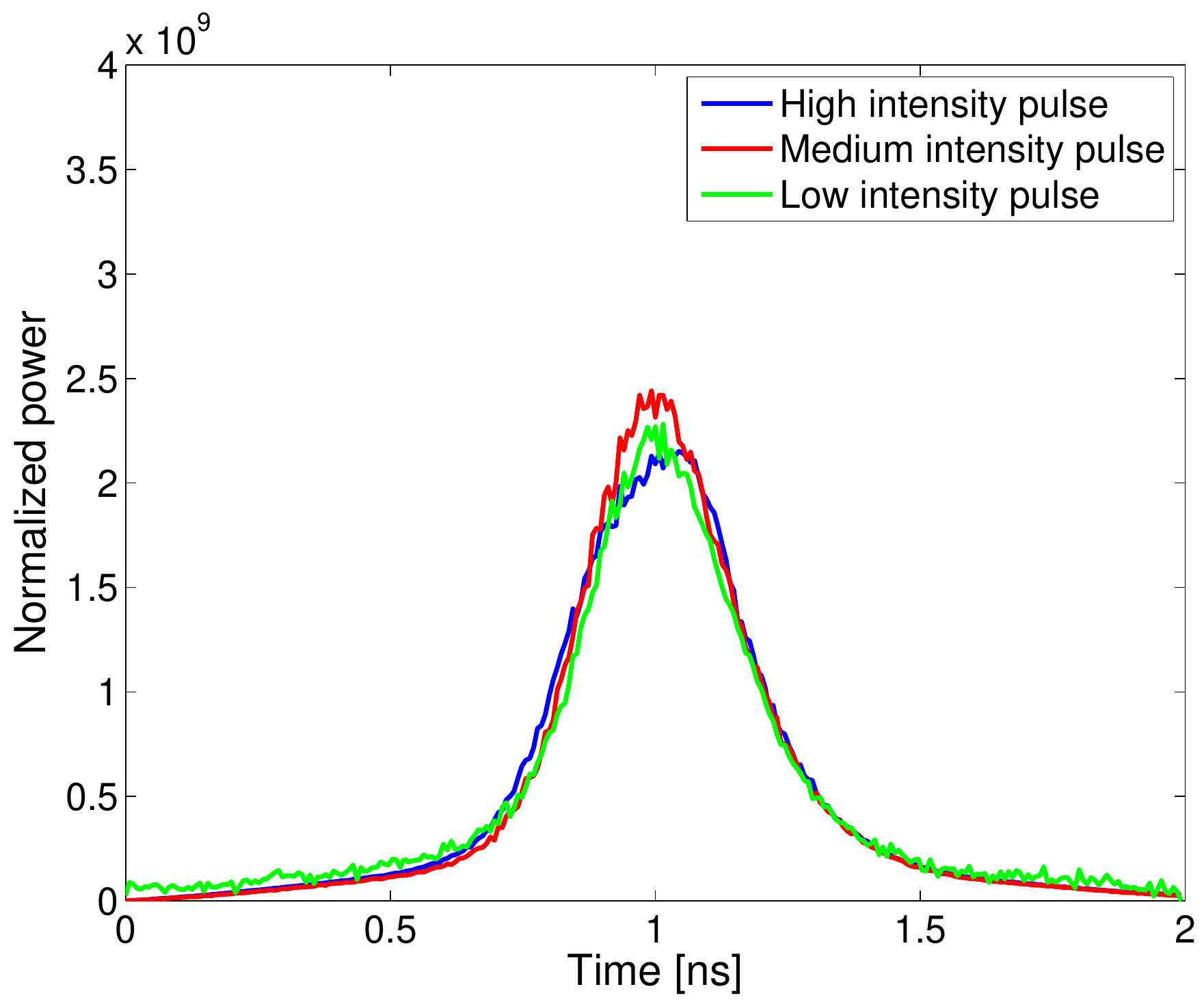}}
\subfigure[]{\includegraphics[angle=0,width=0.49\columnwidth]{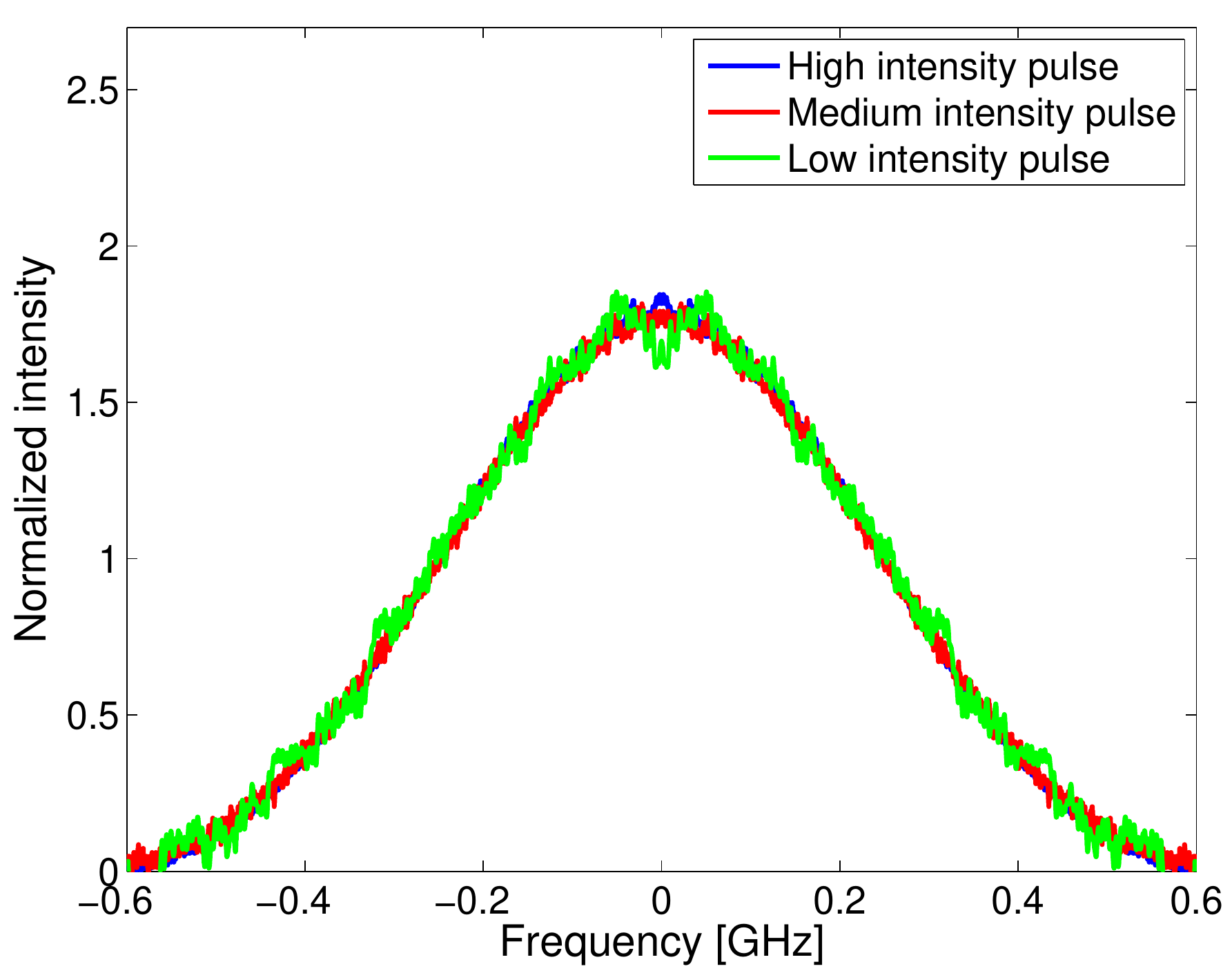}}
\caption{Temporal (a and c) and spectral (b and d) profiles for pulses with three intensity levels. The curves in (c) and (d) have been scaled to allow shape comparison. As expected, these show a high degree of similarity, indicating minimal distortion of the pulses by the amplitude modulator.}
\label{Fig:Intensity_indistinguishability_time_and_spectrum}
\end{center}
\end{figure}

\begin{figure}[htp]
\begin{center}
\subfigure[]{\includegraphics[angle=0,width=0.49\columnwidth]{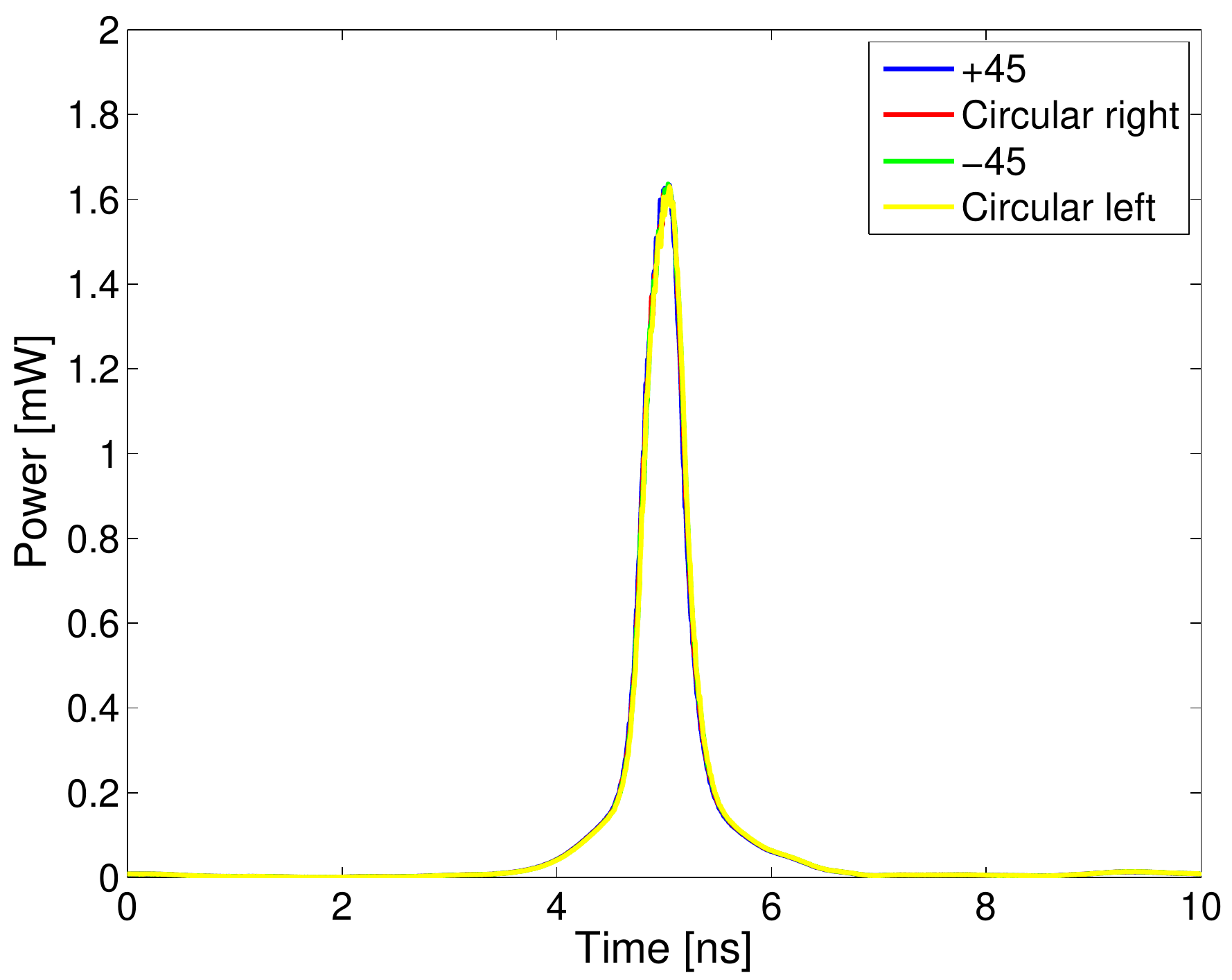}}
\subfigure[]{\includegraphics[angle=0,width=0.49\columnwidth]{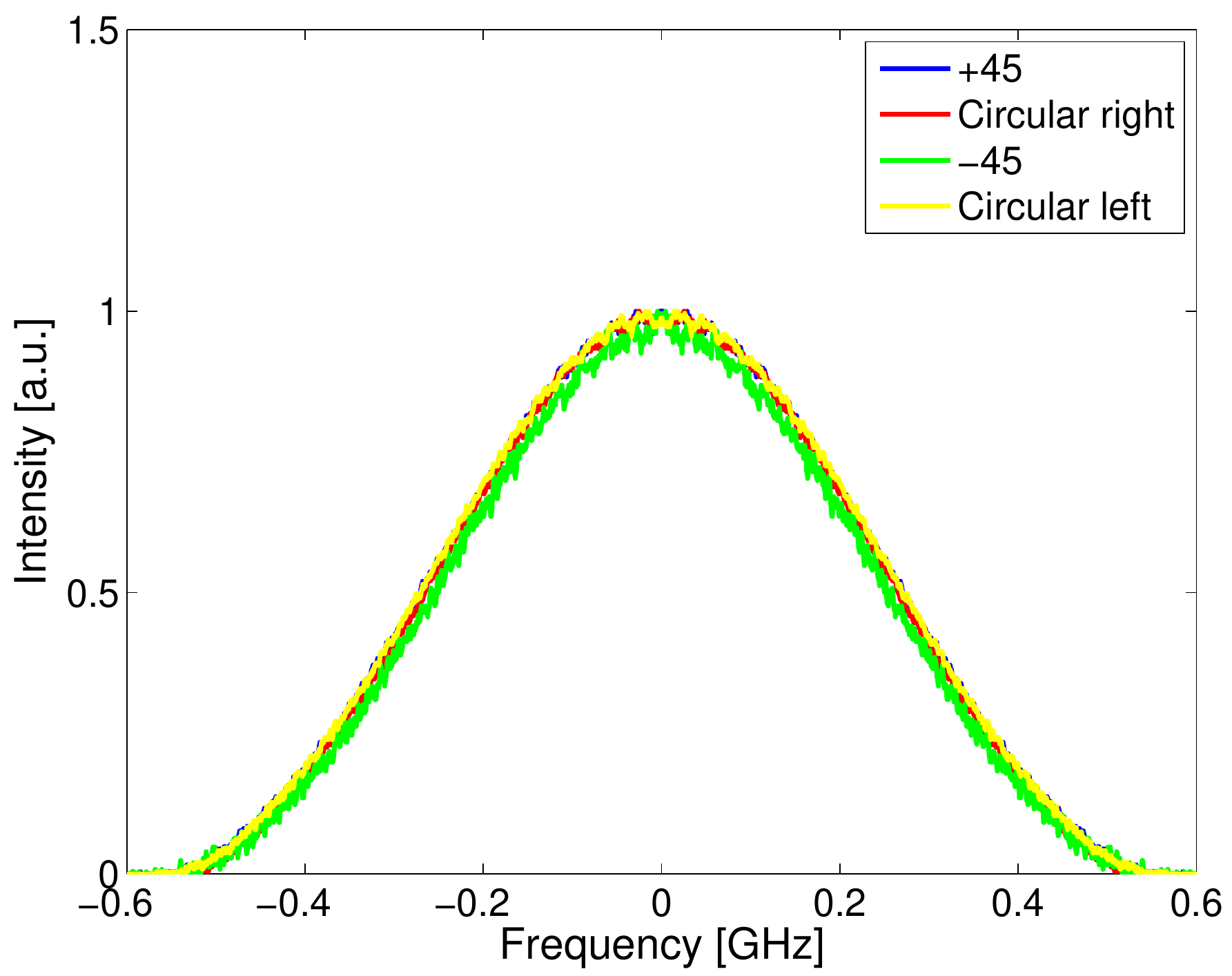}}
\caption{Temporal (a) and spectral (b) profiles for pulses with four polarization states. As expected, these show a high degree of similarity, indicating minimal distortion of the pulses by the polarization modulator.}
\label{Fig:Polarization_indistinguishability_time_and_spectrum}
\end{center}
\end{figure}

Figure \ref{Fig:AM_3_levels_and_PM_Poincare} (a) shows the three different
intensity optical pulses generated after the AM. The attenuations for the medium and low level of
intensity pulses are about 4.65dB and 14.76dB, with respect to the
high intensity pulse. While Figure \ref{Fig:AM_3_levels_and_PM_Poincare} (b) shows the four
polarization states generated after the PM, as measured with a terminating rotating waveplate polarimeter. The RF modulating signal is driven at 100 MHz, in this way, chirp produced at the pulse edges of the RF driving voltage is avoided and intensity and polarization indistinguishability is obtained.
Figure \ref{Fig:Intensity_indistinguishability_time_and_spectrum}
shows pulses with the same polarization but with different
intensity levels with the aim of comparing its temporal and spectral
indistinguishability. A 8 GHz amplified photodiode and a 4 MHz resolution Fabry-Perot interferometer were used for the temporal and spectral measurements, respectively. In order to compare pulses with
different intensity levels, the different pulses are normalized to
their own total intensity. Fig.
\ref{Fig:Polarization_indistinguishability_time_and_spectrum}
shows a similar comparison, but this time pulses have the same
intensity level and different polarization.

\section{QKD performance analysis}
Given the experimental data on the classical optical performance
of the proposed source, a low \textit{Quantum Bit Error Rate}
(QBER) as well as a high Secure Key Rate in the order of Mb/s
are expected. A simulation-based analysis of the expected rates and
performances of a QKD BB84, implementing decoy-state protocol, is
derived below as a demonstration of the potentials for
applications of the proposed FPS.

In a BB84 only single photon pulses contribute to the secure key
while in a 3-state decoy protocol one can obtain a lower bound for
the secure key generation rate as
\begin{equation}
R\geq q \frac{N_\mu}{t}\left\{-Q_\mu f\left(E_\mu\right)H_2\left(E_\mu\right)+Q_1\left[1-H_2\left(e_1\right)\right]\right\}
\end{equation}
where $q$ depends on the implementation (1/2 for the BB84
protocol), $N_{\mu}$ is the total number of detected signal pulses,
$t$ is the time duration of the QKD transmission, $\mu$ represents
the intensity of the signal states, $Q_\mu$ is the gain of the
signal states, $E_\mu$ is the total QBER, $Q_1$ is the gain of
single photon states, $e_1$ is the error rate of single photon
states, $f\left(x\right)$ is the bi-direction error correction
efficiency (taken as $1.16$ \cite{PhysRevA.72.012326}, for an error rate of $1$\%) as a
function of error rate, and $H_2\left(x\right)$ is the binary
Shannon information function, given by
\begin{equation}
H_2\left(x\right)=-x\log_2\left(x\right)-\left(1-x\right)\log_2\left(1-x\right)
\end{equation}

\begin{figure}[htp]
\centering\includegraphics[width=0.8\columnwidth]{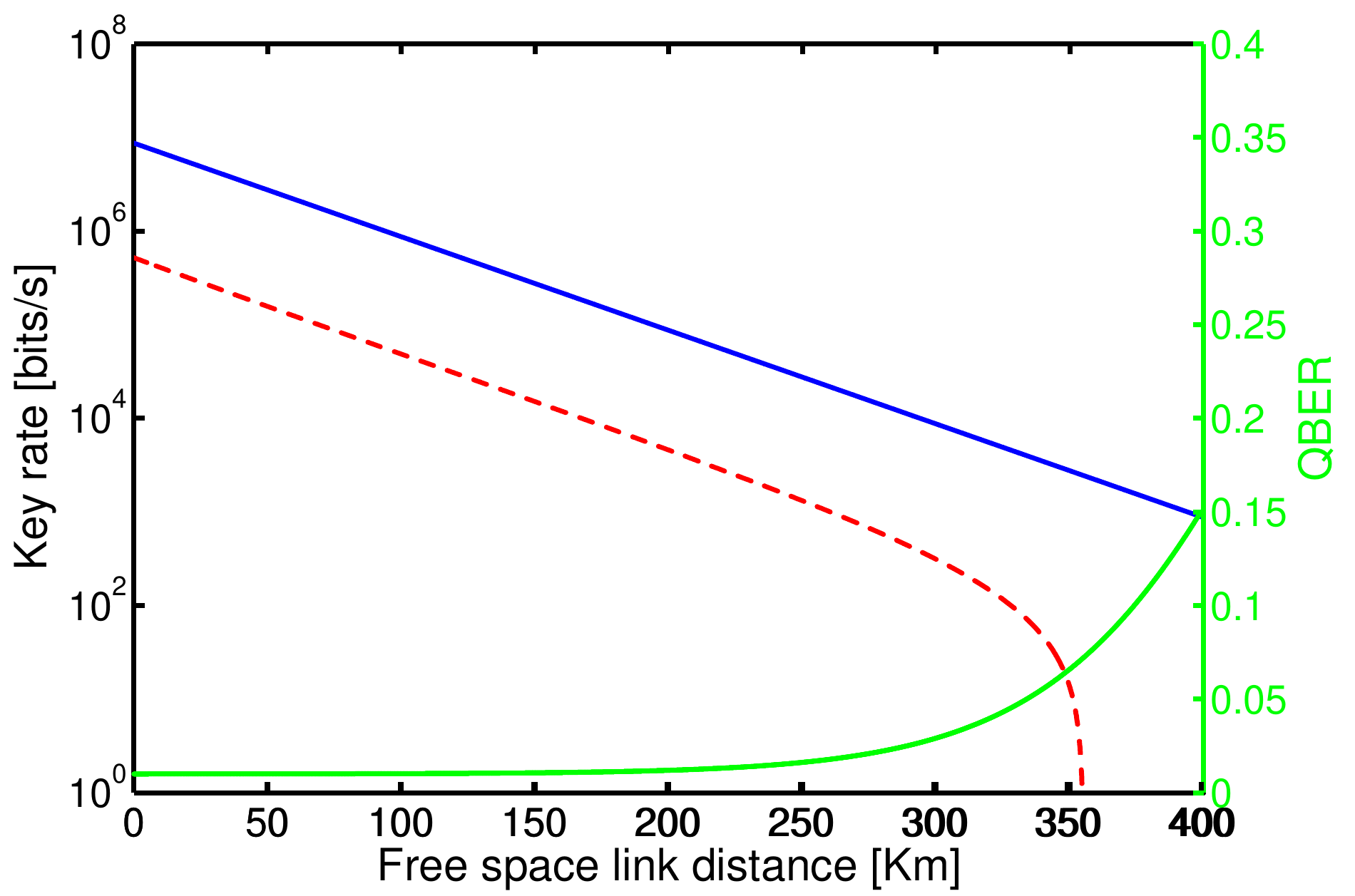}
\caption{QKD BB84 implementing decoy-states simulation
results. Raw Key Rate (blue solid line), Secure Key Rate (red
dashed line) and QBER (green dotted line). In the simulation the
detectors efficiency was set to 50$\%$, free space loss 0.1dB/Km,
5dB were accounted for the loss due to the transmitting and
receiving optical systems, background yield $1\times 10^{-5}$ and
detector misalignment error of $1\%$. All the parameters used in
the simulation are consistent with experimental values reported
in \cite{PhysRevA.72.012326}.} \label{Fig:QKD_simulation_result}
\end{figure}

Figure \ref{Fig:QKD_simulation_result} shows the free space link
distance dependence of the Raw Key Rate, Secure Key Rate and QBER.
The same parameters used for the 20 Km experiment are used for all
the distances considered.

\section{Results and discussion}
Table \ref{Tab:Relevant_parameters_AM_levels} summarizes the
characteristics of the driving RF and corresponding optical pulses
for the three levels of intensity, suitable for a decoy-state
protocol. We believe that, should they be needed, larger intensity
attenuation could be achieved by improved DC voltage bias of the
AM. The AM driving RF signal and the corresponding AM output
quality largely demonstrate the 100 MHz and even beyond capability
of the source. The modulator "ON" window has a duration of at
least 5 ns, much larger than that of the optical pulse. Therefore,
only the amplitude of the optical pulse changes, while the
temporal and spectral shape remain unaltered. In addition low
driving voltages are needed, making the design suitable for
electronic integration with low electrical power consumption
drivers.
\begin{table}
\centering
\caption{Relevant parameters of three generated pulse intensity levels with the amplitude modulator. The RF driving voltages needed are below 1V, which are suitable to be integrated.}
\label{Tab:Relevant_parameters_AM_levels}
\resizebox{\columnwidth}{!}{
\begin{tabular}{l c  c}
\multicolumn{3}{c}{}\\
\hline
            Pulse & AM driver RF signal [mV] & Optical attenuation [dB]\\
\hline
            High intensity level & $460$ & $0$ (reference)\\
            Medium intensity level & $745$ & $4.65$\\
            Low intensity level & $920$ & $14.76$\\
\hline
\end{tabular}}
\end{table}

Table \ref{Tab:Relevant_parameters_PM_states} summarizes the RF
voltages driving the PM generating the four orthogonal states. In
the same table, cross \textit{polarization extinction ratio} (PER)
values for the four different polarization states are given. The
PER values obtained ($>$25dB) are significantly higher than those
required for a low QBER (20dB). As for the AM case, low driving
voltages are needed, suitable for integration with low power
consumption and inexpensive electronics.
\begin{table}
\centering \caption{Relevant parameters of four polarization
states generated with the polarization modulator. Again, the RF
driving voltages needed are low (below 1.5V), which are suitable
to be integrated. The obtained polarization extinction ratio (PER)
largely exceeds the value ($>$ 20 dB) needed to achieve a low
QBER.}
\label{Tab:Relevant_parameters_PM_states}
\resizebox{\columnwidth}{!}{
\begin{tabular}{l c c}
\multicolumn{3}{c}{}\\
\hline
            Polarization & PM driver RF signal [V] &  PER [dB]\\
\hline
            $+45$\textdegree & $0$ & $25.66$\\
            $-45$\textdegree & $1.56$ & $25.84$\\
            Right-handed circular & $0.81$ & $25.65$\\
            Left-handed circular & $-0.76$ & $25.10$\\
\hline
\end{tabular}}
\end{table}

As expected, Figure
\ref{Fig:Intensity_indistinguishability_time_and_spectrum} and
\ref{Fig:Polarization_indistinguishability_time_and_spectrum} show
the high degree of similarity of the pulses, independently of
their polarization or intensity state, indicating minimal pulse
distortion due to the AM and the PM. It has to be noticed that the
small differences for the different intensity pulses are due to
measurement errors. Nevertheless, as commented in section
\ref{Sec:Description_generated_states} polarization statistical similarity
is more important than intensity statistical similarity.
Furthermore, information on the absolute or relative phase between
pulses is not contained in these four figures. However, by design,
the phase of each pulse varies at random between pulses due to the
fact that, as already mentioned, pulses are generated by taking
continuously the laser diode above and below threshold, as
explained in section \ref{Sec:Description_generated_states}.

In the simulation the detectors efficiency was set to 50$\%$, free
space loss 0.1dB/Km, 5dB were accounted for the loss due to the
transmitting and receiving optical systems, background yield $1\times 10^{-5}$ and detector misalignment error of $1\%$. Note that the background yield $Y_0$ includes the detector dark count and other background contributions from stray light, including scattered light from timing pulses \cite{PhysRevA.72.012326,Schmitt-Manderbach2007}, being for larger distances the major cause of secure key rate drop. The
parameters, derived from the values presented above, and results
for the simulated BB84 transmission, implementing the decoy-state
protocol as well as for free space distance of 20 Km, are shown in
Table \ref{Tab:Relevant_parameters_QKD_simulation}. The simulation
has been completed using values taken from
\cite{PhysRevA.72.012326,Dixon:08}, achieving a theoretical Secure Key Rate
of 559.80 Kb/s, which is consistent with the free space achieved
value of 50 Kb/s (over 480 m) reported for a 10 MHz source in \cite{Weier2006}
taking into account that the presented source emits pulses with a
repetition rate one order of magnitude larger.
\begin{table}
\centering \caption{Summary of the QKD simulation parameters
and results for a 20 Km BB84 transmission, implementing the
decoy-state protocol using the experimental data for the FPS. The computed values are for a 20 Km free space distance, where $\mu$, $\nu_1$ and $\nu_2$ are the signal, decoy 1 and decoy 2 (ideally vacuum) states, which have been presented in the previous section, with rates of $0.85$:$0.10$:$0.05$, respectively. The computed values are the gains for the signal $Q_\mu$, decoy 1 $Q_{\nu_1}$ and decoy 2 $Q_{\nu_2}$ states. The QBER for the signal states $e_\mu$, the gain and QBER for the single photon pulses, $Q_1$ and $e_1$, respectively. Finally the lower bound of the secure key rate $R_{secure}$, for the presented source, turns out to be $559.80$ Kb/s.}
\label{Tab:Relevant_parameters_QKD_simulation}
\resizebox{\columnwidth}{!}{
\begin{tabular}{l l c l l}
\multicolumn{5}{c}{}\\
\hline
            Parameter & Value & \vline & Parameter & Value\\
\hline
            Free space link length & $20$ Km & \vline & $Q_\mu$ & $4.87\times 10^{-2}$\\
            $\mu$ & $0.5$ & \vline & $Q_{v_1}$ & $1.70\times 10^{-2}$\\
            $\nu_1$ & $0.125$ & \vline & $Q_{v_2}$ & $1.68\times 10^{-3}$\\
            $\nu_2$ & $0.0167$ & \vline & $e_\mu$ & $1.01\%$\\
            Prob($\mu$:$\nu_1$:$\nu_2$) & $0.85$:$0.10$:$0.05$ & \vline & $Q_1$ & $3.47\times 10^{-2}$\\
            t & $1$s & \vline & $e_1$ & $1.01\%$\\
            $f\left(E_\mu\right)$ & $1.16$ & \vline & $R_{secure}$ & $559.80$ Kb/s\\
\hline
\end{tabular}}
\end{table}

The laser diode is DC biased at $24$mA presenting a DC resistance of $3\Omega$ accounting for $1.7$mW. In addition, an impedance matching circuit has been designed to $50\Omega$ for RF modulation where the electrical pulses of $50$mA, $1$ns wide, at $100$MHz account for $12.5$mW. The modulators do not have a termination resistor, basically, they present an open-ended transmission line with an equivalent loss resistor and parallel capacitor of $5\Omega$ and $10$pF, respectively. Considering the worst case situation where the source is at maximum modulation speed and using the maximum driving voltages, the power consumption for the AM is $2.7$mW and for the PM is $7.7$mW. Thus, the overall power consumption of the integrated module is potentially very low.

\section{Conclusion}\label{Conclusion}
We have shown that, starting from commercially available and
space-qualifiable components, it is possible to build an
integrated transmitter capable of generating the several intensity
and polarization states required for decoy-state QKD. The
experimental demonstration has been carried out at 850 nm with 100
MHz modulation rates. However, taking into consideration that the
modulators bandwidth can go well beyond 10 GHz and operate also at
other wavelengths (e.g. 1550 nm), the source can be easily
scalable to higher bit rates, the upper limit being probably given
by the laser diode itself, and other transmission systems (e.g.
optical fibers).

Although we believe that the proposed source is of general use in
polarization modulation optical systems, especially free-space
links, we have focused our demonstration in preparation for a QKD
experiment using decoy-state protocol, where the indistinguishability of
the pulses, both in the frequency and time domain, is the key for the
security of the link. Given the relatively low driving voltages of
the modulators, the proposed transmitter is potentially low power
consumption and also highly integrable.

\section*{Acknowledgment}
This work was carried out with the financial support of the
Ministerio de Educacion y Ciencia (Spain) through grants
TEC2007-60185, FIS2007-60179, FIS2008-01051 and Consolider Ingenio
CSD2006-00019.

\bibliographystyle{IEEEtran}

\end{document}